\documentclass[12pt]{article}
\usepackage{amssymb}
\usepackage{graphicx}
\usepackage[cp1251]{inputenc}
\usepackage{rotating}

\begin{document}
 \noindent {\footnotesize\it Astrophysical Bulletin, 2026, Vol. 81}
 \newcommand{\dif}{\textrm{d}}

 \noindent
 \begin{tabular}{llllllllllllllllllllllllllllllllllllllllllllll}
 & & & & & & & & & & & & & & & & & & & & & & & & & & & & & & & & & & & & & &\\\hline\hline
 \end{tabular}

  \vskip 0.5cm
  \bigskip
 \bigskip
  \centerline {\large  A STAUDY OF THE SPATIAL EVOLUTION OF THE RADCLIFFE}
  \centerline {\large  WAVE IN A SAMPLE OF YOUNG OPEN STAR CLUSTERS}
   \bigskip
\centerline{ V. V. Bobylev$^1$\footnote [1]{e-mail: bob-v-vzz@rambler.ru},
  A. T. Bajkova$^1$, N. R. Ikhsanov$^{1,2}$ }
 \bigskip
 \centerline {\small \it Central Astronomical Observatory of the Russian Academy of Sciences, Pulkovo}
  \centerline {\small \it Institute of Applied Astronomy of the Russian Academy of Sciences, St. Petersburg}
 \bigskip

{\bf Abstract}---A sample of 139 young open star clusters closely associated with the Radcliffe wave is considered. Modeling their spatial distribution and kinematics over a time interval of 30 Myrs ago and 30 Myrs into the future revealed that they exhibit the main properties characteristic of a Radcliffe wave over the past 10--15 Myr. They are distributed on the galactic $XY$ plane as a long and narrow chain inclined to the $Y$ axis, and exhibit a wave-like behavior of their vertical coordinates up to 15 Myr in the past. This behavior of their vertical coordinates will persist over the interval of 15--20 Myr in the future. A new finding is the presence of vertical perturbations with an amplitude of deviation from the galactic symmetry plane of up to 200 pc over the entire time interval considered in the past, up to $-30$ Myr. This result calls into question the possibility of using a scenario in which the initial disturbance of the interstellar medium is assumed to be the Parker instability of the galactic magnetic field.

  \bigskip

 \subsection*{INTRODUCTION}
The Radcliffe wave was originally discovered during a study of the spatial distribution of a large sample of molecular clouds located in the solar neighborhood at distances up to 3 kpc (Alves et al. 2020). It is a relatively narrow (150--200 pc) wave-like chain of clouds elongated into a line with a length of $\sim$\,2.7\,kpc in the galactic plane. The main feature of this structure is its wave-like character in the direction perpendicular to the plane of symmetry of the galactic disk. The maximum value of the wave amplitude in this direction is reached in the immediate vicinity of the Sun and is approximately 150 pc. The position of the Radcliffe wave is closely related to the Local Arm, the closest edge of which it outlines from the Sun (Bobylev et al. 2025b).

Further studies of the Radcliffe wave showed that the wave-like behavior of the vertical coordinates (in the direction perpendicular to the plane of symmetry of the galactic disk) is also manifested in the distribution of various young objects -- interstellar dust (Lallement et al. 2022; Edenhofer et al. 2024), molecular clouds (Zucker et al. 2023), masers and radio stars (Bobylev et al. 2022), T-Tauri stars (Li, Chen 2022), the youngest massive OB stars (Donada, Figueras 2021; Thulasidharan et al. 2022) and young open star clusters (OC, Donada, Figueras 2021). It was found that the vertical velocities of young OCs in the Radcliffe wave also form a wave, in the range of 15~km s$^{-1}$ (Konietzka et al. 2024). Finally, in the same paper, it was noted that the Radcliffe wave drifts radially in the disk plane toward the galactic anticenter at a velocity of approximately 5~km s$^{-1}$, and that there is also a tangential motion of this structure at a low velocity.

To date, the Radcliffe wave remains a unique phenomenon. Attempts to detect similar structures in the Galaxy (Bobylev 2024; Martinez-Medina et al. 2025) have been unsuccessful. One of the reasons for this may be the short lifetime of such formations. In particular, the age of the Radcliffe wave, based on the spatial distribution of OCs belonging to different age groups, does not exceed 30 million years (Bobylev et al. 2025a). Moreover, the results of modeling the evolution of the local interstellar gas presented in the paper by Li et al. (2024) indicate a relatively rapid evolution of the Radcliffe wave structure and its scale, which, due to the gravitational influence of the Galaxy, will double in only 45 million years. This circumstance significantly reduces the possibility of identifying similar structures in the Galaxy.

There is no consensus on the origin of the Radcliffe wave and its possible connection to the influence of the Galaxy's spiral structure. Several possible scenarios for the origin of the Radcliffe wave have been proposed. One of the first hypotheses put forward was the development of a Kelvin-Helmholtz instability in the galactic disk, arising from the difference in rotational velocities between the dark matter halo and the disk (Fleck 2020). If this scenario were to materialize, Radcliffe wave analogs would be expected in various parts of the disk, which has not yet been observed. The most frequently discussed hypothesis involves the impact of an external impactor on the galactic disk, which has been proposed to be a dwarf galaxy satellite of the Milky Way, a massive clump of dark matter, or a globular cluster (Thulasidharan et al. 2022). The influence of shock waves from several supernova explosions and stellar wind on the structure of the galactic disk, leading to the formation of the Local Bubble or the North Polar Spur, was also discussed (Marchal et al. 2023; Konietzka et al. 2024). The twist angle values for the Orion Arm (which acts as a segment of the spiral arm), compiled from literature data (Bobylev et al. 2025a), lie in the range from $-10^\circ$ to $-13^\circ$. These values differ significantly from the tilt of the molecular cloud chain of $-30^\circ$, which is used to detect the Radcliffe wave. Thus, the connection between the Radcliffe wave and the spiral structure of the Galaxy is not obvious.

The role of the galactic magnetic field in the formation of such an extended structure remains unclear. Reconstruction of the magnetic field geometry in the Radcliffe wave region indicates that the field lines are aligned with the gas-dust medium, following its geometry and forming a significant inclination to the plane of the galactic disk in the solar neighborhood (Panopoulou et al. 2025). One possible scenario for the emergence of such a structure is the development of a Parker instability in the galactic disk, caused by the emergence of magnetic field arches. In its basic characteristics, such as wavelength and development time, this instability corresponds to the parameters of a Radcliffe wave (Bobylev et al. 2025b). However, the root cause of the observed magnetic field disturbance remains elusive. Indeed, in addition to the possible development of instability, one cannot rule out the possibility that the magnetic field structure was deformed as a result of strong disturbance of the gas-dust medium by the external factors listed above. It is not surprising that perturbation of a medium initially frozen in a magnetic field could lead to a distortion of its configuration. This scenario is supported by the fact that a significant portion of young stars form at the crest of the Radcliffe wave, i.e., in a region located at a significant vertical distance from the plane of symmetry of the galactic disk. Explaining this fact within the classical scenario of star formation due to Parker instability is difficult, since the most probable region of star formation in this case is the plane of symmetry of the galactic disk, toward which interstellar gas flows as magnetic field loops emerge (Kaplan and Pikelner 1970). A basis for choosing the most probable of the two magnetic field perturbation scenarios described above can be based on a study of the spatial and kinematic evolution of the Radcliffe wave on a time scale of tens of millions of years. Resolving this pressing problem within the framework of research into the dynamics of molecular clouds is hampered by the lack of sufficient information on the proper motions of these wave components. A much more favorable situation, however, is emerging for open clusters, for which the necessary set of measurement data---trigonometric parallaxes, proper motions, radial velocities, and age estimates---is currently available.

The aim of this paper is to study the spatial evolution of the Radcliffe wave. To do this, we use a sample of open star clusters, constructing and analyzing their galactic orbits for 30 million years into the past and 30 million years into the future. We place particular emphasis on studying the youngest OCs, whose spatial and kinematic properties are as close as possible to those of the molecular clouds in which they formed, and whose initial perturbations led to the formation of the Radcliffe wave structure observed at the present epoch.

\subsection*{METHOD}
In our work, we use two rectangular coordinate systems, heliocentric $(x,y,z)$ and galactocentric $(X,Y,Z).$ In the heliocentric system, the $x$ axis is directed towards the galactic center, the $y$ axis~--- towards the galactic rotation, and the $z$ axis~--- towards the north galactic pole. In the galactocentric system, the $X$ axis is directed towards the Sun from the galactic center, the $Y$ axis~--- towards the galactic rotation, and the $Z$ axis~--- towards the north galactic pole. The distance from the Sun to the galactic rotation axis $R_0$ is taken to be $8.1\pm0.1$~kpc according to the review by Bobylev and Bajkova~(2021), where it was derived as a weighted average of a large number of individual estimates.

\begin{figure}[t]
{ \begin{center}
  \includegraphics[width=0.95\textwidth]{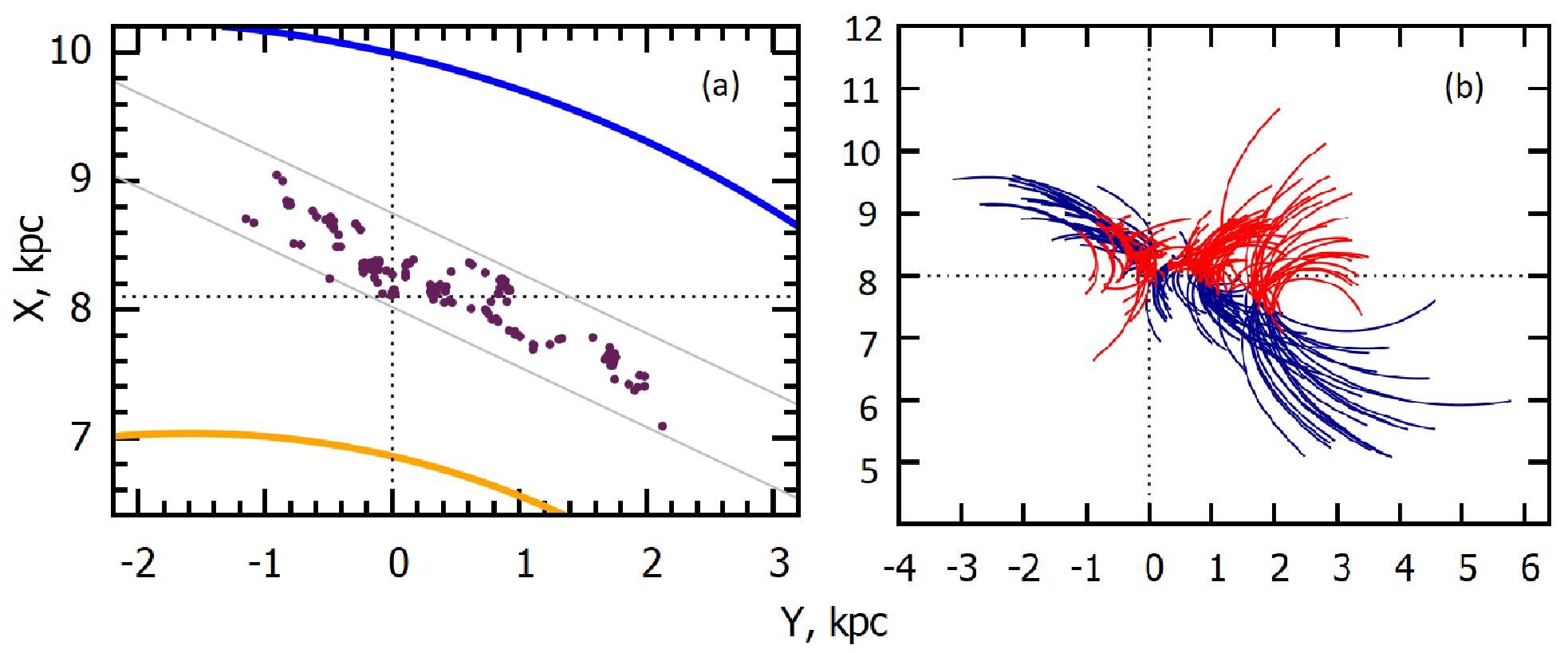}
  \caption{
Distribution of 139 OCs younger than 10~Myr projected onto the galactic $XY $plane (a), where segments of the spiral arms closest to the Sun are indicated by blue (Perseus) and orange (Carina-Sagittarius) lines, thin gray lines denote the boundaries of the selection region; trajectories of 139 OCs in a moving reference frame (relative to the local standard of rest), plotted backward in time over an interval of 30 Myr~--- red lines and forward in time by 30 Myr~--- dark blue lines~(b).
 }
 \label{f1-XY-139}
\end{center}}
\end{figure}
\begin{figure}[t]
{ \begin{center}
  \includegraphics[width=0.55\textwidth]{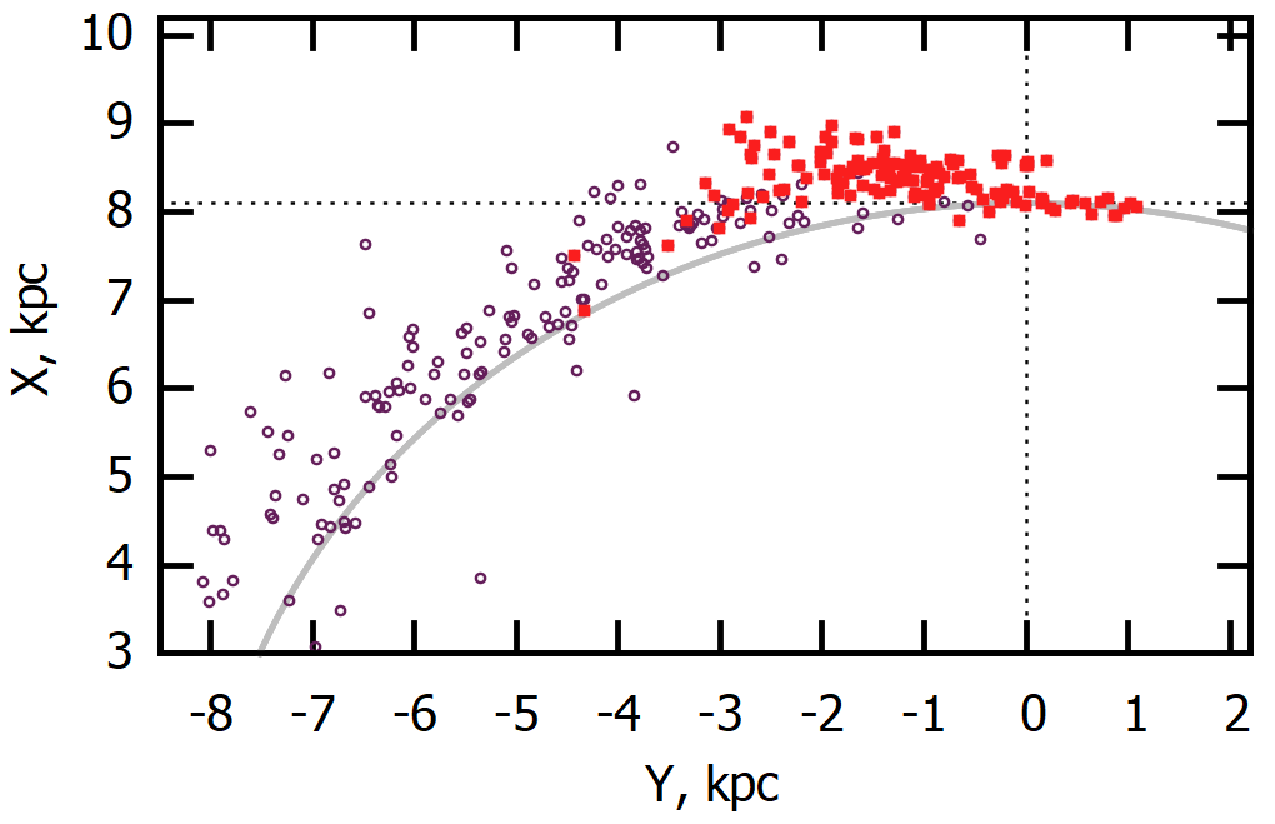}
  \caption{
The birthplaces of OCs that are less than 50 Myrs old are shown by dark circles; OCs younger than 10 Myrs are marked by red squares; the Solar Circle is shown by a grey line.
 }
 \label{f2-XY}
\end{center}}
\end{figure}
\begin{figure}[t]
{ \begin{center}
  \includegraphics[width=0.5\textwidth]{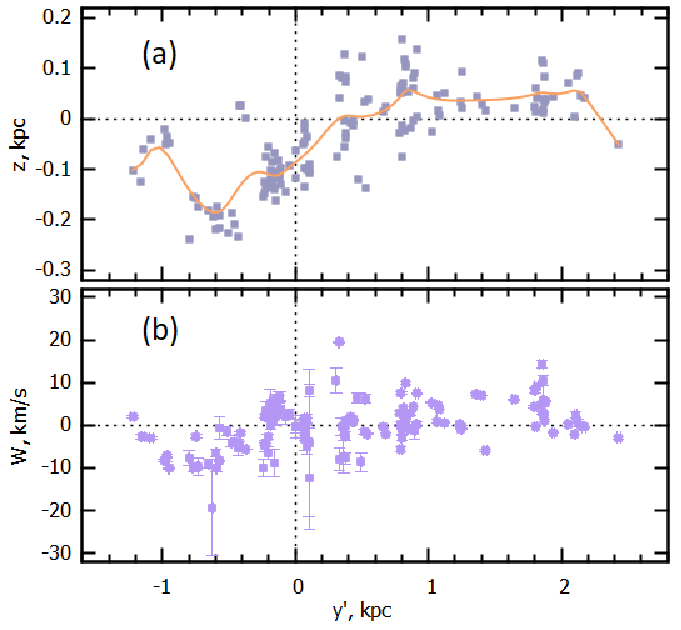}
  \caption{
Currently observed z-coordinates of a sample of OCs younger than 10~Myr (a) and their vertical velocities $W$ as a function of the 
$y'$~coordinate (b).
 }
 \label{f3-139}
\end{center}}
\end{figure}
\begin{figure}[p]
{ \begin{center}
  \includegraphics[width=0.55\textwidth]{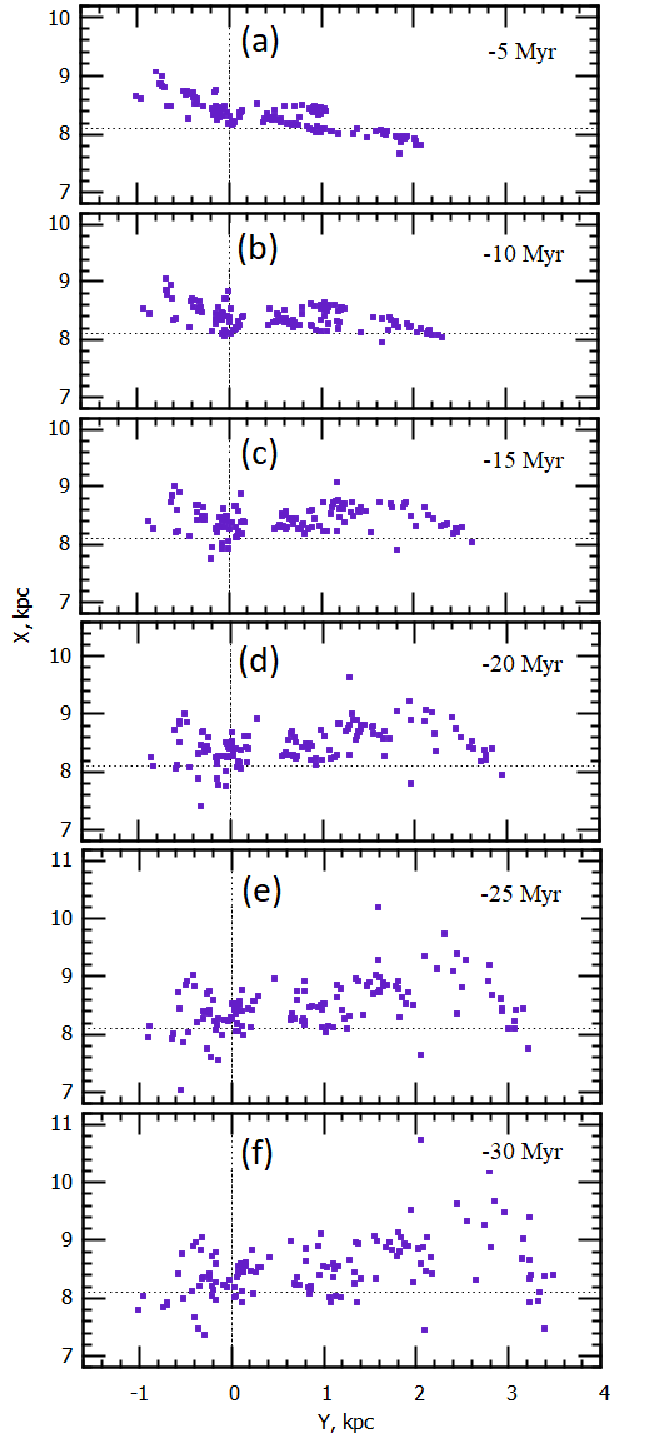}
  \caption{
Distribution of a sample of 139 OCs younger than 10~Myr projected onto the galactic plane $X,Y$ at different times in the past, the local standard of rest is located at the point with coordinates $(X,Y)=(8.1,0)$~kpc.
 }
 \label{f4-Minus}
\end{center}}
\end{figure}
\begin{figure}[p]
{ \begin{center}
  \includegraphics[width=0.85\textwidth]{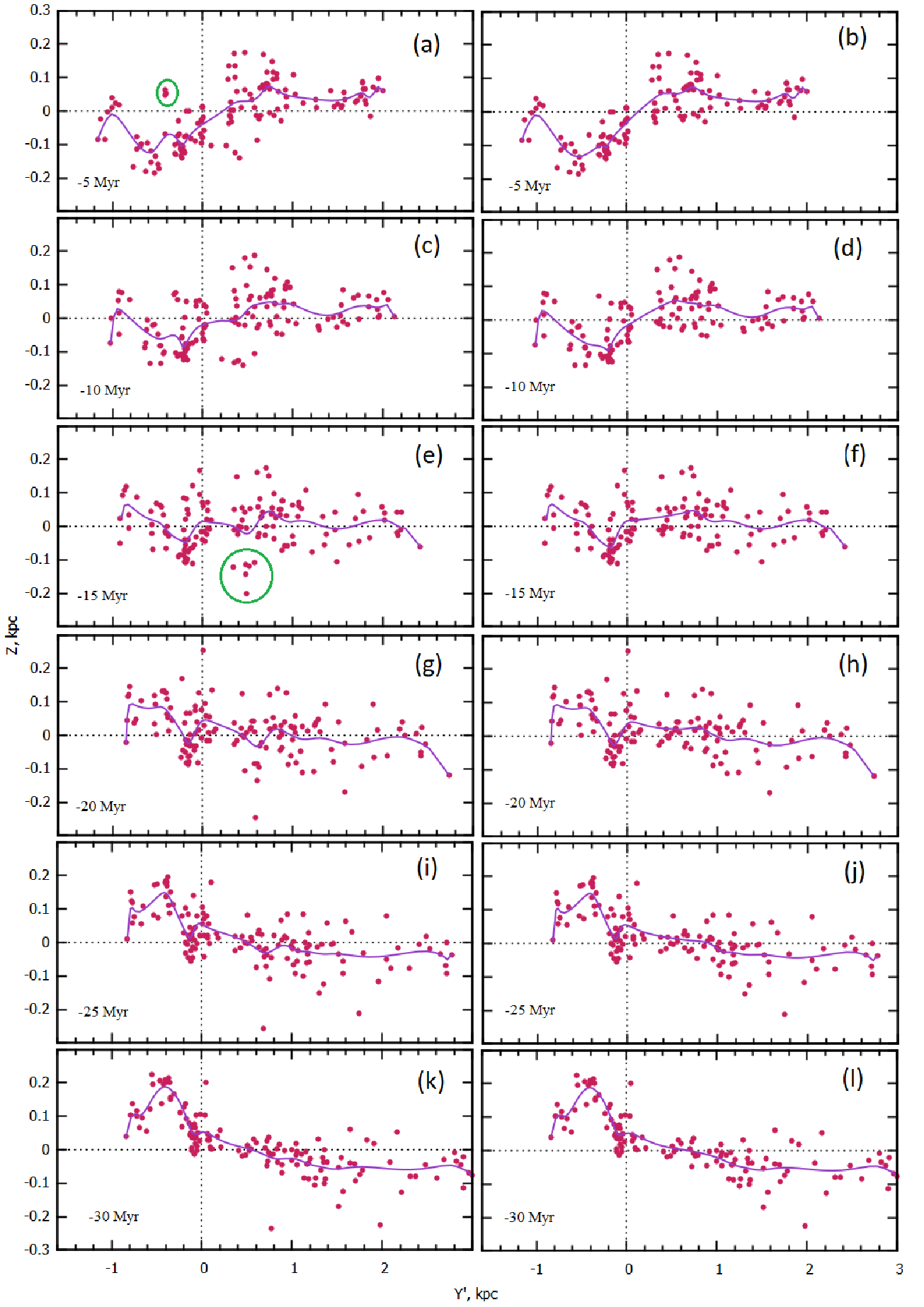}
  \caption{
The z-coordinates of a sample of 139 OCs younger than 10~Myr as a function of the $y'$ coordinate at different times in the past are on the left, and 130 OCs after excluding rebounds (rebounds are marked with green circles in the two left panels) are on the right.
 }
 \label{f5-Myr}
\end{center}}
\end{figure}
\begin{figure}[p]
{ \begin{center}
  \includegraphics[width=0.5\textwidth]{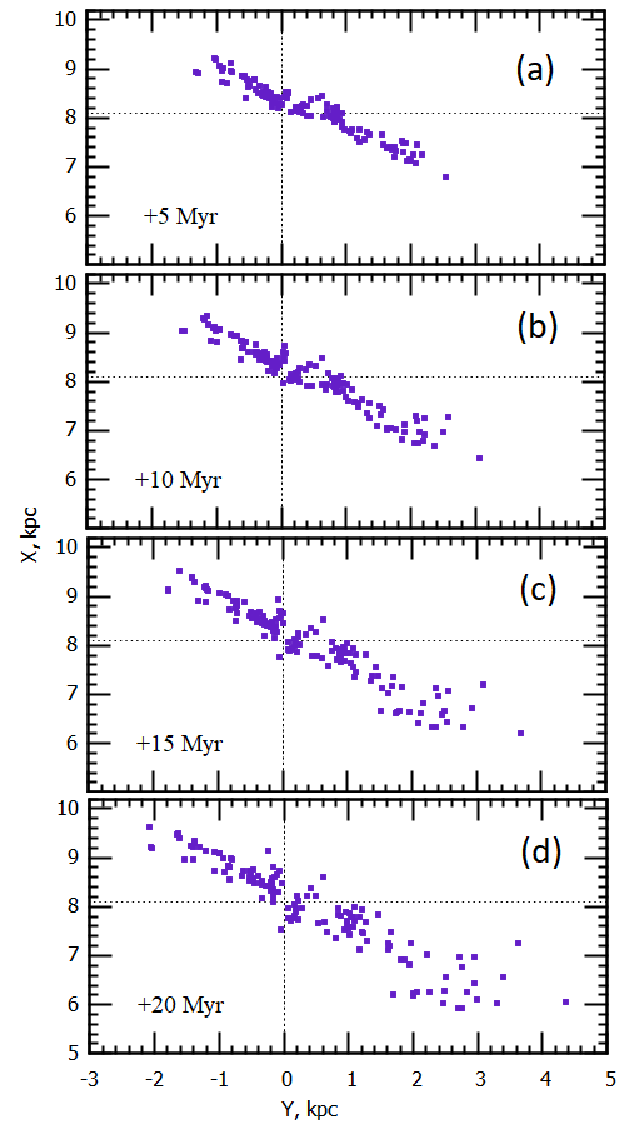}
  \caption{
Distribution of a sample of 139 OCs younger than 10~Myr projected onto the galactic plane $X,Y$ at different times in the future, with the local standard of rest at the point with coordinates $(X,Y)=(8.1,0)$~kpc.
 }
 \label{f6-plus}
\end{center}}
\end{figure}
\begin{figure}[p]
{ \begin{center}
  \includegraphics[width=0.55\textwidth]{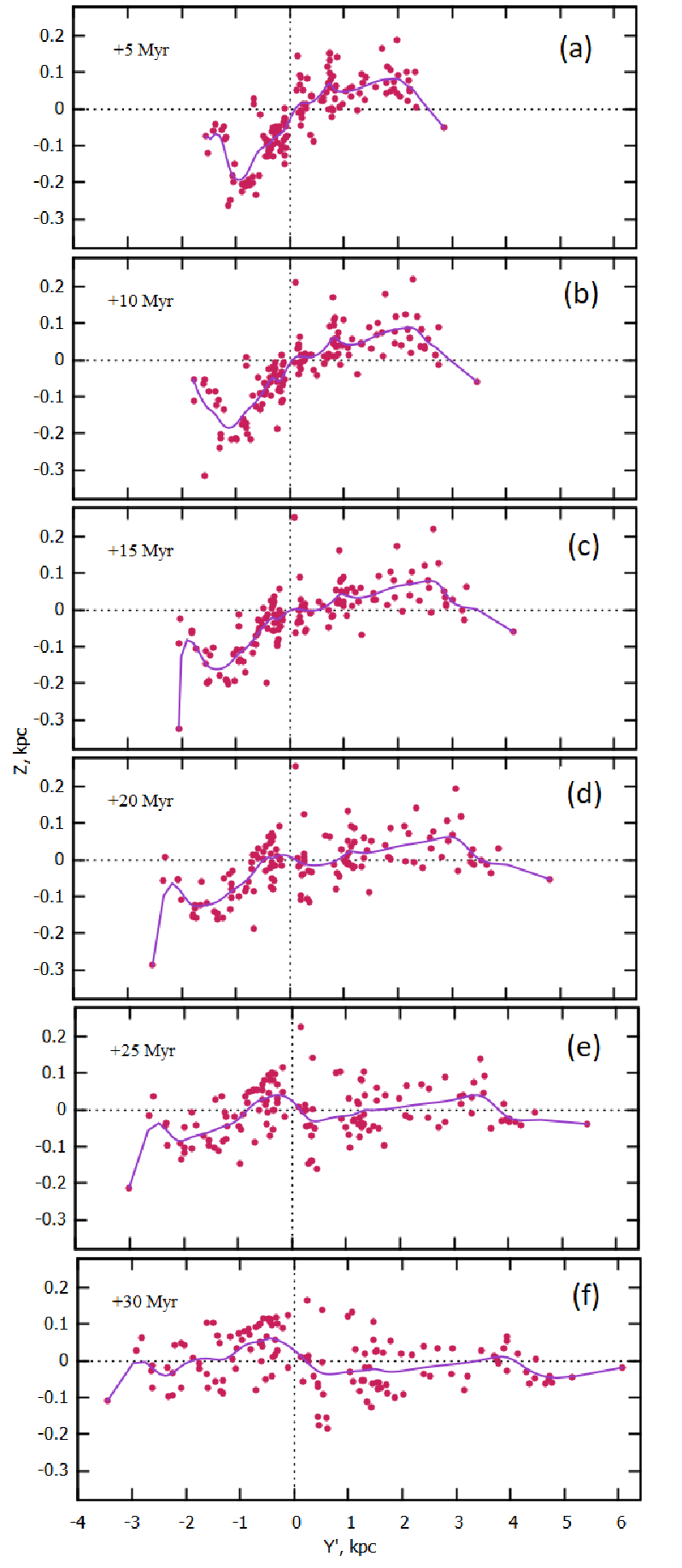}
  \caption{
The $z$ coordinates of a sample of 139 OCs younger than 10~Myr as a function of the $y'$ distance at different times in the future.
 }
 \label{f7-plus}
\end{center}}
\end{figure}
\begin{figure}[t]
{ \begin{center}
  \includegraphics[width=0.75\textwidth]{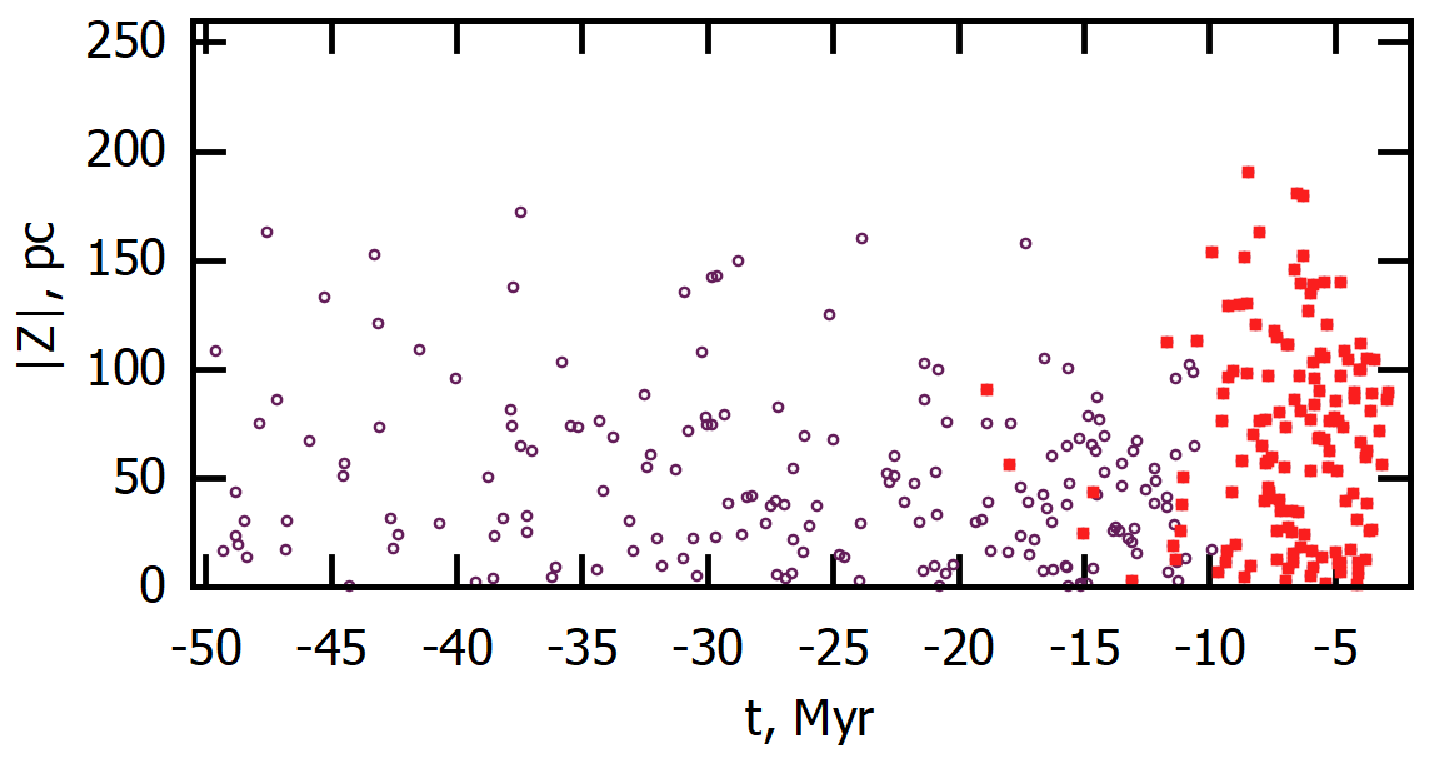}
  \caption{
The $z$-modules of a sample of 139 OCs younger than 10~Myr are red squares and OCs younger than 30 Myr depending on their birth time.
 }
 \label{f8-minus}
\end{center}}
\end{figure}

To construct the galactic orbits of the open clusters, a cylindrical coordinate system ($R,\theta,Z$) with the origin at the galactic center is used. The orbits are calculated using the axisymmetric gravitational potential of the Galaxy $\Phi$, consisting of a central spherical bulge $\Phi_b$, a disk $\Phi_d$, and a massive spherical dark matter halo $\Phi_h=\Phi_b+\Phi_d+\Phi_h.$ The bulge $\Phi_b$ is represented by the Plummer potential (Plummer 1911), the disk $\Phi_d$ is in the form proposed by Miyamoto and Nagai~(1975), and the halo component~--- according to Navarro et al.~(1997). The values of the parameters of the potential model used are given in the paper Bobylev and Bajkova~(2016a), where it is denoted as model~III.

Galactic orbits of open-cluster stars are constructed in a coordinate system related to the local standard of rest. Therefore, the Sun's peculiar velocity relative to the local standard of rest, with values of $(U,V,W)_\odot=(11.0,12.0,7.2)$~km s$^{-1}$ from Sch\"onrich et al. (2010), is excluded from the initial velocities. The Sun's elevation above the galactic plane of symmetry, $h_\odot=16$~pc (Bobylev and Bajkova 2016b), is also taken into account.

Note that open-cluster star orbits can be represented either in a fixed coordinate system $(X,Y,Z)$ or in a system rotating with the local standard of rest.

 \subsection*{DATA}
This paper uses the characteristics of open-cluster clusters from the updated version of the Hunt and Reffert (2024) catalog. This version of the catalog includes 5,647 open star clusters and 1,309 moving groups. Of these, 3,530 open star clusters and 539 moving groups are of high quality, with most located at heliocentric distances less than 4 kpc. The catalog contains the average values of the open cluster trigonometric parallaxes, proper motions, and radial velocities calculated using data from the Gaia DR3 catalog (Gaia Collab. 2022). Cluster age estimates obtained by these authors using the isochrone method are also presented.

The publication by Hunt and Reffert (2024) is primarily devoted to dynamical estimates of OCs, such as their masses and radii, as well as to studying the nature of their connectivity. A detailed description of the derivation of average parallaxes, proper motions, radial velocities, and ages is given in Hunt and Reffert (2023). It is interesting to note that the relative errors in the average parallaxes and proper motions of OCs obtained by these authors are small and do not exceed 10--15\%. Random errors in the average radial velocities of OCs do not exceed 10~km s$^{-1}$ for a significant portion of the catalog, although there are clusters with significantly higher levels of such errors. The age estimates of OCs were obtained with relative errors of 30--50\%.

We selected the OCs, probably associated with the Radcliffe wave, in the zone bounded by two lines on the plane $xy:$
\begin{equation}
 \begin{array}{lll}
 x= y\tan\beta-0.65, \\
 x= y\tan\beta+0.05,
 \end{array}
 \label{xy-0.65}
\end{equation}
where the angle $\beta=25^\circ$ was chosen to optimally manifest the wave. The center of the selection zone does not pass through the origin, but intersects the $x$-axis at $x_1$, so the transition to the $y'$-axis was performed using the formula
\begin{equation}
y'= y\cos\beta+(x-x_1)\sin\beta,
\label{y'-dx}
\end{equation}
where $x_1=-0.28$~kpc.

\subsection*{RESULTS}
We selected open star clusters younger than 10~Myr, with the index ``o'' (open cluster) in the Hunt and Reffert~(2024) catalog, with parallaxes, proper motions, and radial velocities. After selection according to the boundary conditions~(\ref{xy-0.65}), the sample included 139~OCs with an average age of 6.2~Myr.

Fig. ~\ref{f1-XY-139}(a) shows the distribution of the 139 selected young open star clusters projected onto the galactic $XY$ plane at the present time. The figure shows that the selection zone is quite wide, but most of the open star clusters are elongated. This distribution is very similar to the distribution of molecular clouds from which the Radcliffe wave was discovered by Alves et al. (2020). This suggests that the kinematic properties of the selected OCs correspond to those of their parent molecular clouds, whose kinematic modeling provides reliable information on the spatial evolution of the Radcliffe wave.

Fig.~\ref{f1-XY-139}(b) shows the trajectories of 139 young OCs in a moving reference frame, plotted for 30~Myrs forward and 30~Myrs back, including the evolution of the parent molecular clouds up to the moment of star formation.

In the same selection region shown in Fig.~\ref{f1-XY-139}(a), 337 OCs with ages less than 50~Myrs were selected. Their birthplaces were determined by constructing galactic orbits backward in time, by a time interval corresponding to their age. The birthplaces of such OCs are presented in Fig.~\ref{f2-XY}, where OCs younger than 10 Myrs are marked with red squares. As can be seen from Fig.~\ref{f1-XY-139}(a) and Fig.~\ref{f2-XY}, most OCs younger than 10 Myrs are currently located very close to their birthplace and generally retain a structure inclined to the $Y$-axis. That is, this sample of OCs is fairly homogeneous in a kinematic sense, although it does contain a small fraction of ``runners''.

Figure~\ref{f3-139} shows the dependences of the vertical coordinates $z$ and velocities $W$ on the coordinate $y'$ for 139 young OCs. The orange line in Figure~\ref{f3-139}(a) marks the averaged coordinate values (the line was drawn when plotting the figure in the gnuplot program). The distribution of points in Figure~\ref{f3-139}(a) has a wave-shaped form, which is consistent with the results of OC analysis obtained by other authors (e.g., Donada, Figueras~2021; Bobylev et al. 2025a; Konietzka et al. 2024). The distribution of vertical velocities in Figure~\ref{f3-139}(b) also has a wave-shaped form.

Fig.~\ref{f4-Minus} shows the distribution of a sample of 139 young OCs projected onto the galactic plane $X,Y$ at different moments in the past in a rotating coordinate system relative to the local standard of rest.

Fig. ~\ref{f5-Myr} shows the dependence of the vertical coordinates $z$ on the $y'$ coordinate for the selected OCs at different points in the past. The left panels of the figure (a, c, e, g, i, k) show such dependences for all 139 OCs. The averaged curves are also presented. Significant deviations from the averaged values (bounces) are marked with green circles in panels (a) and (e). A total of 9 such bouncing OCs were detected. The right panels of the figure (b, d, f, h, j, l) show the dependences without taking into account the bounces, using 130 OCs. Comparing the curves in panels (a) and (b), (c) and (d), the elimination of local deviations from the fairly smooth curves is most clearly evident in panels (e) and (f).

Fig.~\ref{f6-plus} shows the distribution of the 139 selected young OCs projected onto the galactic $X,Y$ plane relative to the future local standard of rest. To save space, only four time slices are shown, as the overall trend is obvious.

Fig.~\ref{f7-plus} shows the vertical coordinates of 139 young OCs as functions of their distance $y'$at various points in the future.

Fig.~\ref{f8-minus} shows the absolute values of the $z$ coordinates for both a sample of 139 OCs younger than 10 Myr old and a larger sample of OCs younger than 50 Myr old (from the same selection zone) as functions of their birth times. This figure was constructed, in particular, to test the possible influence of the Parker instability of the galactic magnetic field on the formation of wave-like irregularities in the galactic disk. One prediction of this scenario is the preferential localization of star-forming regions near the plane of symmetry of the galactic disk. As can be seen from Fig.~\ref{f8-minus}, the birth regions of 139 OCs younger than 10~Myrs, which make up the Radcliffe wave, are located, however, at a significant distance from the plane of symmetry of the galactic disk, far from objects that do not belong to this structure. This indicates that either the Parker instability is not the primary cause of the perturbation of the galactic magnetic field in the Radcliffe wave, or the emerging magnetic field has a more complex geometry, precluding the possibility of interstellar gas flowing along the field lines of the emerging magnetic field toward the plane of symmetry of the galaxy and the formation of initial centers of star formation in these regions.

\subsection*{DISCUSSION}
The sample of 139 OSCs under consideration, with an average age of 6.2 million years, exhibits the key properties characteristic of a Radcliffe wave over the past 10---15 Myrs. First, over this time period, they are spatially distributed as a fairly long and narrow chain, inclined to the $Y$-axis. Second, they exhibit a wave-like behavior of their vertical coordinates up to 15 Myrs in the past.
In other words, the parent clouds from which these OCs formed exhibit, over a significant time interval, properties currently observed in the chain of molecular clouds from which the Radcliffe wave was first discovered. It is also interesting to note that the wave-like behavior of the vertical coordinates of this sample will persist for 15-20 Myrs in the future. This means that there is a mechanism, as yet unknown, that maintains the vertical oscillations observed in the Radcliffe wave over a time interval of about 30 million years ($\pm 15$\,million years from the present).

Our findings of a) significant stretching of the 139~OC over time in the future and b) further destruction of this structure under the influence of differential rotation of the Galaxy and the galactic tide in 20-25 million years are in good agreement with the results of gas cloud kinematics modeling performed in Li et al. (2024). Note that the chain of our OCs in the $X,Y$plane has a wave-like character both in the past (Fig.~\ref{f4-Minus}(a,b,c)) and in the future (Fig.~\ref{f6-plus}(a,b)).

The most interesting of the new effects we discovered is the presence of significant-amplitude vertical disturbances throughout the previous epoch of Radcliffe wave evolution, lasting up to 30~Myrs (see Fig.~\ref{f5-Myr}). The amplitude of such disturbances reaches $z\sim200$~pc and the general appearance of the structure differs significantly from that characteristic of the Radcliffe wave (Fig.~\ref{f5-Myr}(b,d,f)).

\subsection*{CONCLUSION}
Using data from the Hunt and Reffert (2024) catalog of open star clusters, this paper generated a sample of potentially Radcliffe-wave-related open star clusters with ages no older than 10\,million years from a selection zone inclined to the galactic $Y$-axis by an angle of $25^\circ$. The sample included 139 clusters with an average age of 6.2\,million years.

We modeled their spatial distribution and kinematics over a time interval of 30 million years. We found that the OCs we selected for analysis exhibit the main properties characteristic of a Radcliffe wave over the past 10--15 million years. Specifically, they are distributed across the galactic $X,Y$ plane as a long, narrow wave-like chain with a significant inclination to the $y$-axis and a significant amplitude of vertical deviations. Our modeling shows that this behavior of their vertical coordinates will persist over an interval of 15--20 million years in the future.

A fundamentally new result is the significant amplitude (up to $\sim$200\,pc from the galactic symmetry plane) of the initial vertical perturbations of the Radcliffe wave components throughout its evolution over the past 30~million years. The formation region of a significant fraction of the young stars belonging to this structure was initially located at a significant vertical distance from the galactic disk's symmetry plane. In light of this result, a scenario based on the initial perturbation of the interstellar medium by the Parker instability of the magnetic field proves ineffective or requires significant modification.

\medskip
The authors are grateful to the reviewer for helpful comments that contributed to the improvement of this paper.

 \bigskip\medskip  {BIBLIOGRAPHY}\medskip{\small \begin{enumerate}

 \item
J. Alves, C. Zucker, A.A. Goodman, et al., Nature {\bf 578}, 237 (2020).

 \item
V.V. Bobylev, A.T. Bajkova, and Yu.N. Mishurov, Astron. Lett. {\bf 48}, 434 (2022).

\item
V.V. Bobylev, A.T. Bajkova, Astron. Lett. {\bf 42}, 567 (2016a).

 \item
V.V. Bobylev, A.T. Bajkova, Astron. Lett. {\bf 42}, 1 (2016b).

\item
V.V. Bobylev, A.T. Bajkova, Astron. Rep. {\bf 65}, 498 (2021).

\item
V.V. Bobylev, Astron. Lett. {\bf 50}, 796 (2024).

\item
V.V. Bobylev, N.R. Ikhsanov,  A.T. Bajkova, Astrophys. Bull. {\bf 80}, Issue~2, 181 (2025a).

\item
V.V. Bobylev, N.R. Ikhsanov,  A.T. Bajkova, Astron. Rep. {\bf 69}, 786 (2025b).

 \item
J. Donada, F. Figueras, arXiv: 2111.04685 (2021).

\item
G. Edenhofer, C. Zucker, P. Frank, et al., Astron Astrophys. {\bf 685}, A82 (2024).

\item
R. Fleck, Nature {\bf 583}, 24 (2020).

 \item
Gaia Collaboration (A. Vallenari, et al.),  Astron. Astrophys. {\bf 674}, 1 (2023).

\item
E.L. Hunt, S. Reffert, Astron. Astrophys. {\bf 673}, A114, (2023).

\item
E.L. Hunt, S. Reffert, Astron. Astrophys. {\bf 696}, A42, (2024).

\item
S. A. Kaplan and S. B. Pikelner, The Interstellar Medium (Harvard Univ. Press, Harvard, 1970).

 \item
R. Konietzka, A.A. Goodman, C. Zucker, et al.,  Nature {\bf 628}, 62 (2024).

  \item
R. Lallement, J.L. Vergely, C. Babusiaux, et al., Astron. Astrophys. {\bf 661}, 147  (2022).

 \item
G.-X. Li, B.-Q. Chen, MNRAS {\bf 517}, L102 (2022).

 \item
G.-X. Li, J.-X. Zhou, and B.-Q. Chen, Research Notes of the AAS {\bf 8}, id. 316 (2024).

 \item
A. Marchal, P.G. Martin, Astrophys. J. {\bf 942}, 70 (2023).

 \item
L. Martinez-Medina, E. Poggio, and E. Moreno-Hilario, MNRAS {\bf 542}, L94 (2025).

 \item
M. Miyamoto,  R. Nagai, PASP {\bf 27}, 533 (1975).

\item
J.F. Navarro, C.S. Frenk, and S.D.M. White, Astrophys. J. {\bf 490}, 493 (1997).

\item
G.V. Panopoulou, C. Zucker, D. Clemens, V. Pelgrims, et~al., Astron. Astrophys. 694, A97 (2025).

\item
H.C. Plummer, MNRAS {\bf 71}, 460 (1911).

 \item
R. Sch\"onrich, J. Binney, and W. Dehnen, MNRAS {\bf 403}, 1829 (2010).

 \item
L. Thulasidharan, E. D'Onghia, E. Poggio, et al., Astron. Astrophys. {\bf 660}, 12 (2022).

 \item
C. Zucker, J. Alves, A. Goodman, S. Meingast, and P. Galli, Protostars and Planets VII, ASP Conf. Ser., Vol. 534, Proc. conf. held 10--15 April 2023 at Kyoto, Japan. Eds. S.-I. Inutsuka, Y. Aikawa, T. Muto, K. Tomida, and M. Tamura. San Francisco: Astron. Soc. Pacific, p. 43 (2023).

\end{enumerate}
  \end{document}